\pdfoutput=1
\documentclass[%
 reprint,
superscriptaddress,
 amsmath,amssymb,
 aps, float
]{revtex4-2}

\usepackage{graphicx}
\usepackage{dcolumn}
\usepackage{bm, xcolor, siunitx}
\begin{document}

\preprint{APS/123-QED}

\title{Plastic vortex creep and dimensional crossovers in the highly anisotropic superconductor HgBa$_2$CuO$_{4+x}$}

\author{Haley M. Cole}
\affiliation{Department of Physics, Colorado School of Mines, Golden, CO 80401}

\author{Michael B. Venuti}
\affiliation{Department of Physics, Colorado School of Mines, Golden, CO 80401}

\author{Brian Gorman}
\affiliation{Department of Materials and Metallurgical Engineering, Colorado School of Mines, Golden, CO 80401}%

\author{Eric D. Bauer}
\affiliation{Los Alamos National Laboratory, Los Alamos, NM 87545 (USA)}
    
\author{Mun K. Chan}
\affiliation{Pulsed Field Facility, National High Magnetic Field Laboratory, Los Alamos National Laboratory, Los Alamos, NM 87545 (USA)}
  
\author{Serena Eley}
\affiliation{Department of Physics, Colorado School of Mines, Golden, CO 80401}%

\date{\today}

\begin{abstract}

 In type-II superconductors exposed to magnetic fields between upper and lower critical values, $H_{c1}$ and $H_{c2}$, penetrating magnetic flux forms a lattice of vortices whose motion can induce dissipation.  Consequently, the magnetization $M$ of superconductors is typically progressively weakened with increasing magnetic field $B \propto n_v$ (for vortex density $n_v$).  However, some materials exhibit a non-monotonic $M(B)$, presenting a maximum in $M$ at what is known as the second magnetization peak. This phenomenon appears in most classes of superconductors, including low $T_c$ materials, iron-based, and cuprates, complicating pinpointing its origin and garnering intense interest.  Here, we report on vortex dynamics in optimally doped and overdoped HgBa$_2$CuO$_{4+x}$ crystals, with a focus on a regime in which plastic deformations of the vortex lattice govern magnetic properties. Specifically, we find that both crystals exhibit conspicuous second magnetization peaks and, from measurements of the field- and temperature- dependent vortex creep rates, identify and characterize phase boundaries between elastic and plastic vortex dynamics, as well as multiple previously unreported transitions within the plastic flow regime.  We find that the second magnetization peak coincides with the elastic-to-plastic crossover for a very small range of high fields, and a sharp crossover within the plastic flow regime for a wider range of lower fields.  We find evidence that this transition in the plastic flow regime is due to a dimensional crossover, specifically a transition from 3D to 2D plastic dynamics.

\end{abstract}

\maketitle

\section{\label{sec:introduction} Introduction}

The electromagnetic properties of many seemingly disparate condensed matter systems are dictated by the dynamics of inlying elastic media --- including charge density waves in materials with highly anisotropic band structure \cite{Gruner1998}, domain walls in ferroelectrics, skyrmions in magnets with strong spin-orbit coupling \cite{Nii2014}, and vortex matter in type-II superconductors \cite{Eley2021}.  In these systems, competition between disorder, elasticity, thermal energy, and driving forces from currents can engender elastic or plastic deformations and determine phase boundaries between the two regimes.  In superconductors, plastic flow has received considerably less attention than elastic and glassy regimes, despite its technological relevance for e.g. large-scale high-$T_c$ applications that tend to operate at high temperatures ($T/T_c>0.5$) and fields, often within the plastic flow regime.

In the elastic regime, vortices may be elastically deformed while remaining in their equilibrium positions in a quasi-ordered phase, and the system starts to melt when thermal energy surpasses the elastic energy barriers. The energy barrier that pins vortices, $U(J) \propto J^{-\mu}$, grows infinitely with decreasing current density $J$ (signifying a glass state), characterized by the glassy exponent $\mu>0$ \cite{Blatter1994b}.  When $U(J)$ surpasses the plastic energy barrier, plastic flow occurs which qualitatively differs from the elastic regime in that there are different vortex channels with different dynamics. 

In the plastic flow regime, coherent domains of the vortex lattice are separated by dislocations (line defects) and motion can be dislocation-mediated, similar to the diffusion of dislocations in atomic solids \cite{hirth1982, Abulafia1996}. Some domains may be static whereas others may move at different rates relative to each other \cite{Aranson1996, PhysRevLett.76.831}.  These dynamics often likened to that of conglomerates of ice floes, in which each ice floe is analogous to a quasi-ordered domain of vortices.  Vortex motion may then occur through a variety of distinct dynamic arrangements:  For example, channels of moving vortices can flow between stationary vortex lattice islands \cite{Jensen1989, Koshelev1994, Jensen1990} or large quasi-ordered domains (ice floe) can slide with respect to each other.\cite{Abulafia1996}

One way to determine the relevant energy barriers, and ultimately assess whether the dynamics are elastic or plastic, is to study the rate of thermal activation over these barriers --- the vortex creep rate.  Collective creep theory predicts that, in the elastic regime, $U_{el}(B,J) = U_0(B)(J_c/J)^\mu \propto B^\nu J^{-\mu}$ for magnetic field $B$, critical current density $J_c$, and positive critical exponents $\nu$ and $\mu$, whereas in the plastic flow regime the energy barrier $U_{pl}$ is non-diverging with increasing $J$ \cite{Abulafia1996, Feigelman1989, Blatter1994b}. The creep process is determined by the lowest of the two energy barriers: consequently, a crossover to the plastic flow regime is expected when $U_{pl}$ becomes less than $U_{el}$, generally at low $J$ or high $B$.

Here, we report on a systematic study of vortex dynamics in HgBa$_2$CuO$_{4+x}$ (Hg1201) single crystals, ideal testbeds for studying the effects of thermal fluctuations on 3D and 2D vortex dynamics as well as dimensional crossovers, owing to high anisotropy and high critical temperature $T_c$.  Hg1201 is also of broad interest due to the presence of charge–density–wave correlations that cause quantum oscillations at low temperatures \cite{Chan9782, Barisic2013}. Using extensive magnetization studies, we map the appearance of a second magnetization peak at $H_{smp}(T)$, a crossover between elastic and plastic dynamics at $H_{ep}(T)$, and crossovers within the plastic flow regime suggesting multiple distinct dynamic arrangements. Notably, we find that the second magnetization peak coincides with the elastic-to-plastic crossover at high magnetic fields and low temperatures and a crossover within the plastic flow regime at intermediate fields and temperatures.   It may also correspond with a dimensional crossover from 3D to 2D plastic dynamics. 


\section{Results and Discussion}

\begin{figure}[h]
\includegraphics[scale=0.45, trim=0 0 255 0, clip]{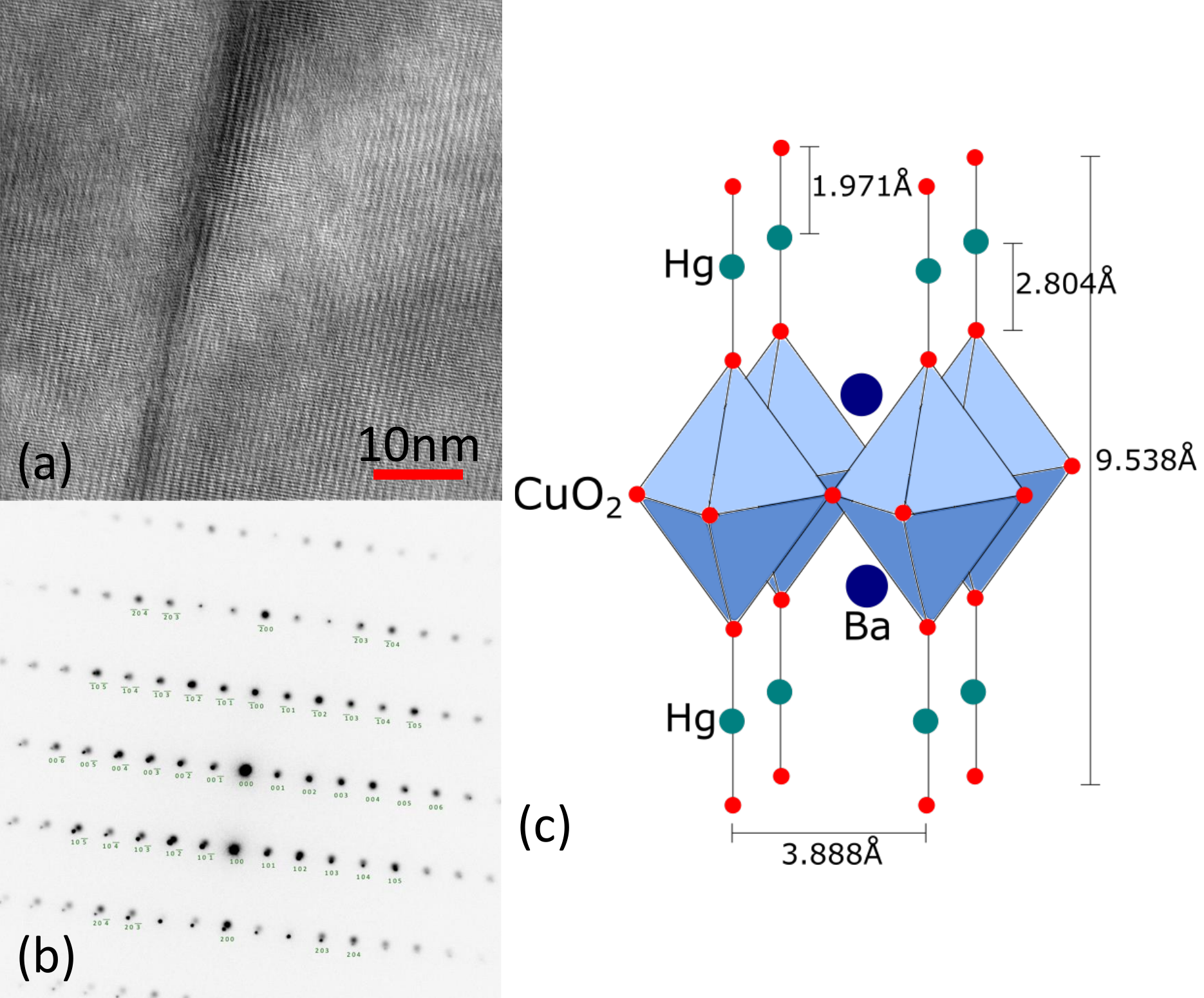}
\caption{\label{fig:FigTEM} (a) High Resolution Bright Field transmission electron micrograph of a planar defect in an optimally-doped Hg1201 single crystal.  (b) Selected Area Electron Diffraction pattern overlaid with a simulated pattern using a [001] zone axis.  The planar defect is observed to exist on the (100) plane of the tetragonal structure shown in (c). (c) Crystal structure of Hg1201, featuring tetragonal symmetry in the space group P4/mmm and a single CuO$_2$ plane per primitive cell \cite{Tabis2014}.}
\end{figure}

We grew, analyzed the microstructure of, and performed magnetization studies on an optimally doped and an overdoped Hg1201 single crystal. Details on the growth method are provided in the Methods section. Transmission electron microscopy studies of an optimally doped crystal, shown in Fig. \ref{fig:FigTEM}, reveal a clean microstructure, with only dislocations evident within the resolution of our system.  Accordingly, our crystals are absent of defects that could exert strong pinning forces on vortices (such as twin boundaries or large precipitates), such that we may expect vortex pinning to primarily originate from weak collective effects from point disorder and that the effects of anisotropy may strongly influence magnetic properties.  

Magnetization measurements $M(T,H,t)$ were performed using a Quantum Design superconducting quantum interference device (SQUID) magnetometer, in which the magnetic field was aligned with the sample's c-axis ($ H \parallel c$), $T$ indicates temperature, and $t$ is time. By measuring $M$ versus $T$ at $5 \textnormal{ Oe}$, after zero-field cooling, we find that the onset critical temperatures $T_c$ of the optimally and overdoped crystals are $ \SI{95.9}{\kelvin}$ and $\SI{90.0}{\kelvin}$, respectively, which is consistent with previous work~\cite{Barisic2008, Pelloquin1997, Villard1999, Wagner1993, Viallet1997}.

\begin{figure}[h]
\centering
\includegraphics[height=2.2in]{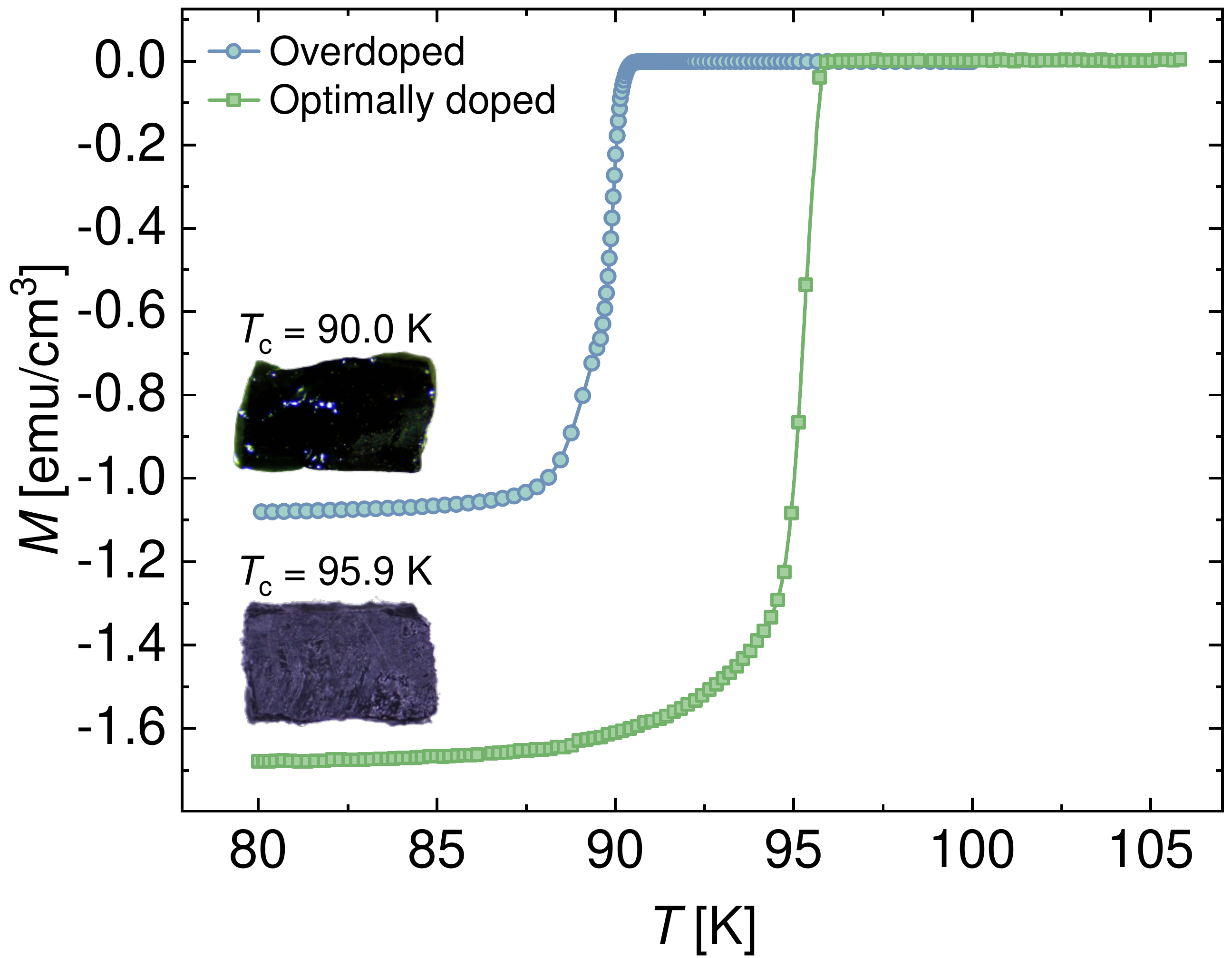}
\caption{\label{fig:FigTc} Temperature-dependent magnetization $M(T)$ measured at $\mu_0H=5~\textnormal{Oe}$ after zero-field cooling, revealing $T_c = \SI{95.9}{\kelvin}$ and $T_c = \SI{90.0}{\kelvin}$, for our optimally doped and overdoped samples, respectively. The insets display optical images of the samples.}
\end{figure}

\begin{figure}[h]
\centering
\includegraphics[height=2.2in]{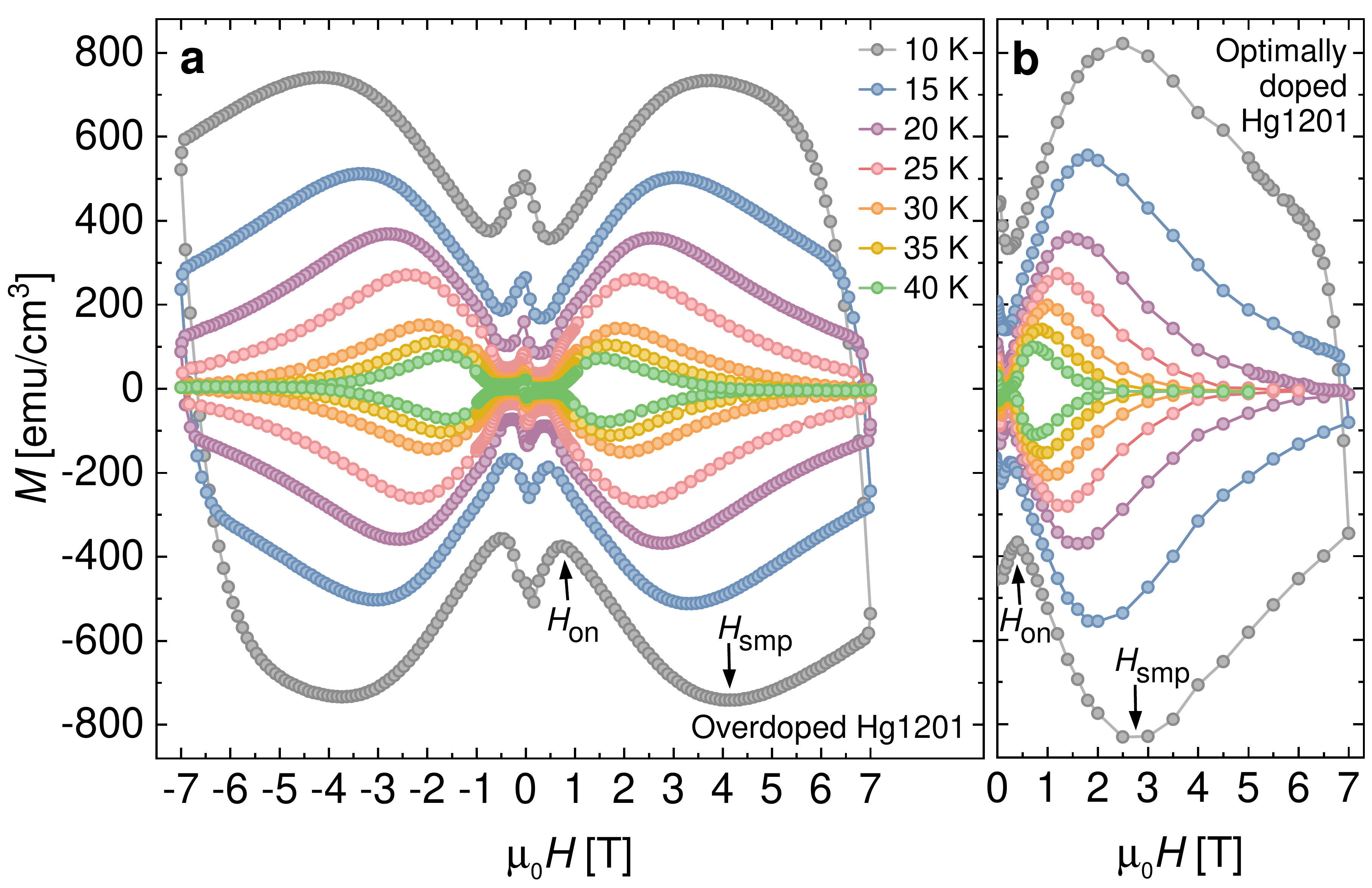}
\caption{\label{fig:FigMvsH} Isothermal magnetic hysteresis loops $M(H)$ at select temperatures for our (a) overdoped and (b) optimally doped Hg1201 single crystals. Each curve exhibits a second magnetization peak at $H_{smp}$ and low-field dip at $H_{on}$.}
\end{figure}

Figure \ref{fig:FigMvsH} shows the isothermal field-dependent magnetization $M(H)$ curves collected for both samples. The curves exhibit a conspicuous peak at $H_{smp}$ that is known as the second magnetization peak (SMP). Second magnetization peaks have been reported in studies of most classes of superconductors, including low-$T_c$ \cite{Banerjee2000, Peng1994}, iron-based \cite{Miu2012, Fang2011, Zhou2016a, Salem-Sugui2010, Pramanik2011a, Shen2010, Ionescu2018, PhysRevB.78.224506, PhysRevB.85.134532, PhysRevB.80.012510, Polichetti2021}, and highly anisotropic \cite{Chowdhury2003, Konczykowski2000} materials and other cuprates \cite{Boudissa2006a, Ionescu2018, Miu2001, PhysRevB.50.7016, KATAYAMA2003382} as well as previous work on Hg1201 \cite{Eley2020, Daignere2000, Villard1999, Pissas1999, Pissas1998, Stamopoulos2002}.  This peak is typically thought to originate from a crossover between vortex pinning regimes, such as elastic-to-plastic transitions~\cite{Miu2012, Zhou2016a, Salem-Sugui2010, Shen2010}, structural phase transitions \cite{Gilardi2002, Rosenstein2005, PhysRevB.81.092504, Miu2020} in the vortex lattice, or dimensional crossovers.
In a previous study~\cite{Eley2020} of an optimally doped Hg1201 crystal, we found that the elastic-to-plastic crossover is not responsible for the SMP in Hg1201.  Though we did note a correspondence between the SMP and the elastic-to-plastic crossover at low temperatures $T/T_c<0.2$, this overlap disappears at higher temperatures, in which the SMP occurred within the plastic flow regime. In this work, we compare the field and temperature at which the second magnetization peak occurs to the elastic-to-plastic crossover, then proceed to characterize the plastic flow regime, which shows evidence of distinct dynamics arrangements that may be responsible for the SMP.  To this end, we extract the elastic and plastic energy barriers, and their dependencies on the current and magnetic field from magnetic relaxation measurements to determine the vortex creep rates.

Vortex creep measurements allow us to probe the vortex structure, dynamics, and interactions (e.g. vortex-vortex, vortex-defect). The vortex creep rate naturally depends on the energy barrier that must be overcome for vortex motion. According to collective creep theories, this energy barrier 
\begin{align}
    U_{act}(J) = (U_p/\mu)\left[(J_{c0}/J)^\mu - 1\right] \label{eq:Uact}
\end{align}
depends on the current critical current in the absence of thermal activation $J_{c0}$, $J=0$ energy barrier $U_p$, and the glassy exponent $\mu$, which is related to the size and dimensionality of the vortex bundle that hops during the creep process \cite{Blatter1994b, Yeshurun1996b}. Specifically, $\mu$ depends on whether dynamics are driven by single vortices or a vortex bundle of lateral dimension $R_c$ smaller than (small bundle), comparable to (medium bundle), or larger than (large bundle) the penetration depth $\lambda_{ab}$.  For example, for 3D vortices, the dynamics of single vortices, small vortex bundles, and large vortex bundles are expected to produce a $\mu$ of $1/7$, $3/2$ or $5/2$, and $7/9$, respectively \cite{Blatter1994b}.
In the case of 2D vortices, $\mu = 7/4, 13/16,$ and $1/2$ is expected for creep of small, medium, and large vortex bundles, respectively \cite{Blatter1994b, Vinokur1995, narlikar2005hts}. By combining Eq. \eqref{eq:Uact} with the creep time $t = t_0e^{U_{act}(J)/k_B T}$ (related to the vortex penetration time \cite{Yeshurun1996b}), we find that the dissipation generated by creep should cause the persistent current to decay over time as $J(t) = J_{c0} [1+(\mu k_B T/U_{p})\ln(t/t_0)]^{-1/\mu}$ and that the thermal vortex creep rate is
\begin{align}
    S \sim \Big| \frac{d \ln J}{d \ln t}  \Big| = \frac{k_B T}{U_p + \mu k_B T \ln(t/t_0)},\label{eq:STHeqn}
\end{align}
where $t_0 \approx 1-10\textnormal{ }\mu\textnormal{s}$ \cite{Yeshurun1996b}.

Because the magnetization in type-II superconductors is proportional to current ($M \propto J$), creep rates can readily be extracted from measurements of $M$ versus time $t$. Likewise, as seen from Eq.~\eqref{eq:STHeqn}, knowledge of $S(T,H)$ provides access to both $U_{p}$ and $\mu$. Consequently, creep measurements are indispensable for revealing the size of the energy barrier, determining its dependence on current, field, and temperature, and also ascertaining whether the dynamics are particle-like, elastic, or plastic.

\begin{figure}[h]
\centering
\includegraphics[width=2.6in]{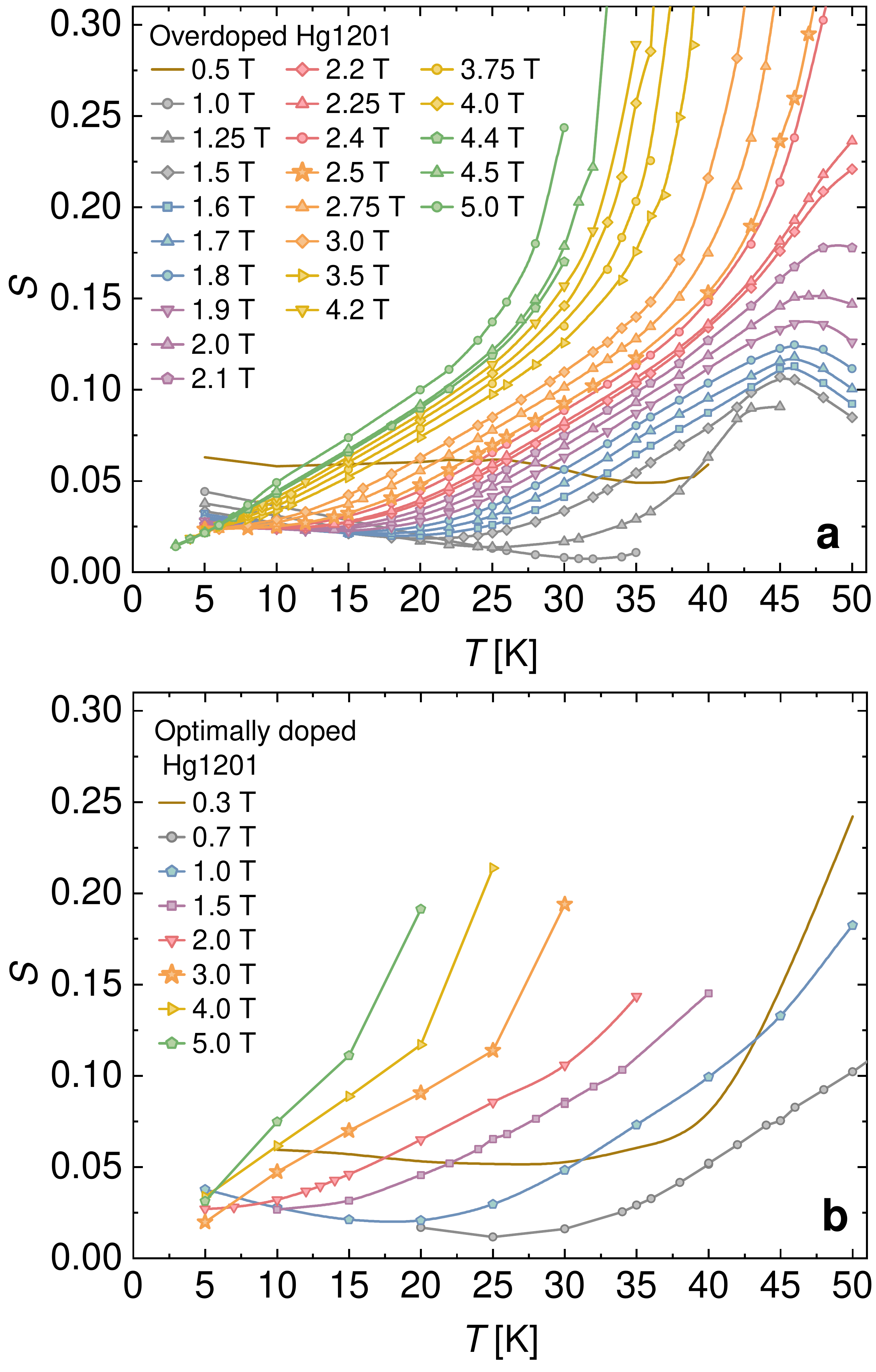}
\caption{\label{fig:FigCreep} Temperature dependence of the vortex creep rates in different applied magnetic fields for \textbf{(a)} the overdoped and \textbf{(b)} the optimally doped Hg1201 crystals.  The error bars are determined from the standard deviations of linear fit to $\log{m} - \log{t}$ (where $m$ is the magnetic moment) and are smaller than the symbol size.}
\end{figure}

Figure \ref{fig:FigCreep}(a) shows the temperature and field dependence of the creep rates in both Hg1201 crystals.  Over most of the temperature and field range, the creep rate increases monotonically with temperature and field, with a few exceptions. First, in both samples, creep at the lowest fields is faster than at higher fields and temperature independent between $\SI{1.8}{\kelvin}$ and $ \SI{40}{\kelvin}$, plateauing around $S \sim 0.06$.  Second, data for the overdoped sample exhibits a small, anomalous peak in $S(T)$ around \SIrange{45}{48}{\kelvin} for fields of \SIrange{1.25}{2.1}{\tesla}. Lastly, in both samples, $S$ decreases with increasing temperature up to $\SI{25}{\kelvin}$ – $\SI{30}{\kelvin}$ then positively correlates with further increases in temperature.  This different dynamics at low fields may be related to the self-field $H_{sf}$ generated by induced currents being higher than the applied field: at fields below $H_{sf}$, vortex lines over a large region of the sample peripheries are quite curved, straightening as the applied field approaches $H_{sf}$.  In these samples, the self-field $\mu_0 H_{sf}=\gamma J_{sf} \delta /\pi$ (where $\delta$ = thickness, $J_{sf}$ is the zero-field critical current, and $\gamma$ is the anisotropy) ranges from \SI{5.4}{\tesla} to \SI{0.4}{\tesla} for temperatures \SIrange{5}{30}{\kelvin} in the overdoped sample.  The low-field dynamics may also be related to the mechanism responsible for the low-field dip at $H_{on}$ (Fig. \ref{fig:FigMvsH}), which is unresolved in many materials. In Ref. \cite{Eley2020}, we found a dip in $S(H)$ near $H_{on}$, which was later shown in Ref. \cite{Polichetti2021} to be a common trend among many materials, including some other highly anisotropic materials (e.g. Bi$_2$Sr$_2$CaCu$_2$O$_{8+x}$), some less anisotropic cuprates (e.g. YBa$_2$Cu$_3$O$_{7-x}$), iron-based superconductors (e.g. BaFe$_2$(As$_{0.72}$P$_{0.28}$)$_2$), and (K, Ba)BiO$_3$ \cite{Polichetti2021}.

The low-field plateau in $S(T)$ is a signature of glassiness over a broad temperature range, as seen in YBa$_2$Cu$_3$O$_{7-x}$ single crystals \cite{Thompson1993b}. From Eq.~\eqref{eq:STHeqn}, we may expect a plateau to occur when $U_p \ll \mu_0 k_B T \ln(t/t_0)$ such that $S \sim [\mu \ln(t/t_0)]^{-1}$ becomes independent of temperature.  For our measurement window of $t \sim$ 1 hour, $\ln(t/t_0) \approx 27$ such that $\mu \sim 0.6$, close to the expectation of 0.5 for creep of large bundles of 2D vortices in both samples \cite{Blatter1994b, Vinokur1995, narlikar2005hts}.

\subsection{Elastic vortex dynamics and the elastic-to-plastic crossover}

Analysis of the current dependence of the effective activation energy can provide direct experimental access to $\mu$, given that $U_p$ is typically otherwise unknown. Combining equations \eqref{eq:Uact} and \eqref{eq:STHeqn}, and considering the exponential for the creep time $t_0$, we find that the effective pinning energy is:
\begin{align}
    U^\star \equiv \frac{T}{S} = U_p\left(\frac{J_{c0}}{J}\right)^\mu. \label{eq:UstarJ}
\end{align}

\begin{figure}[h]
\centering
\includegraphics[width=2.6in]{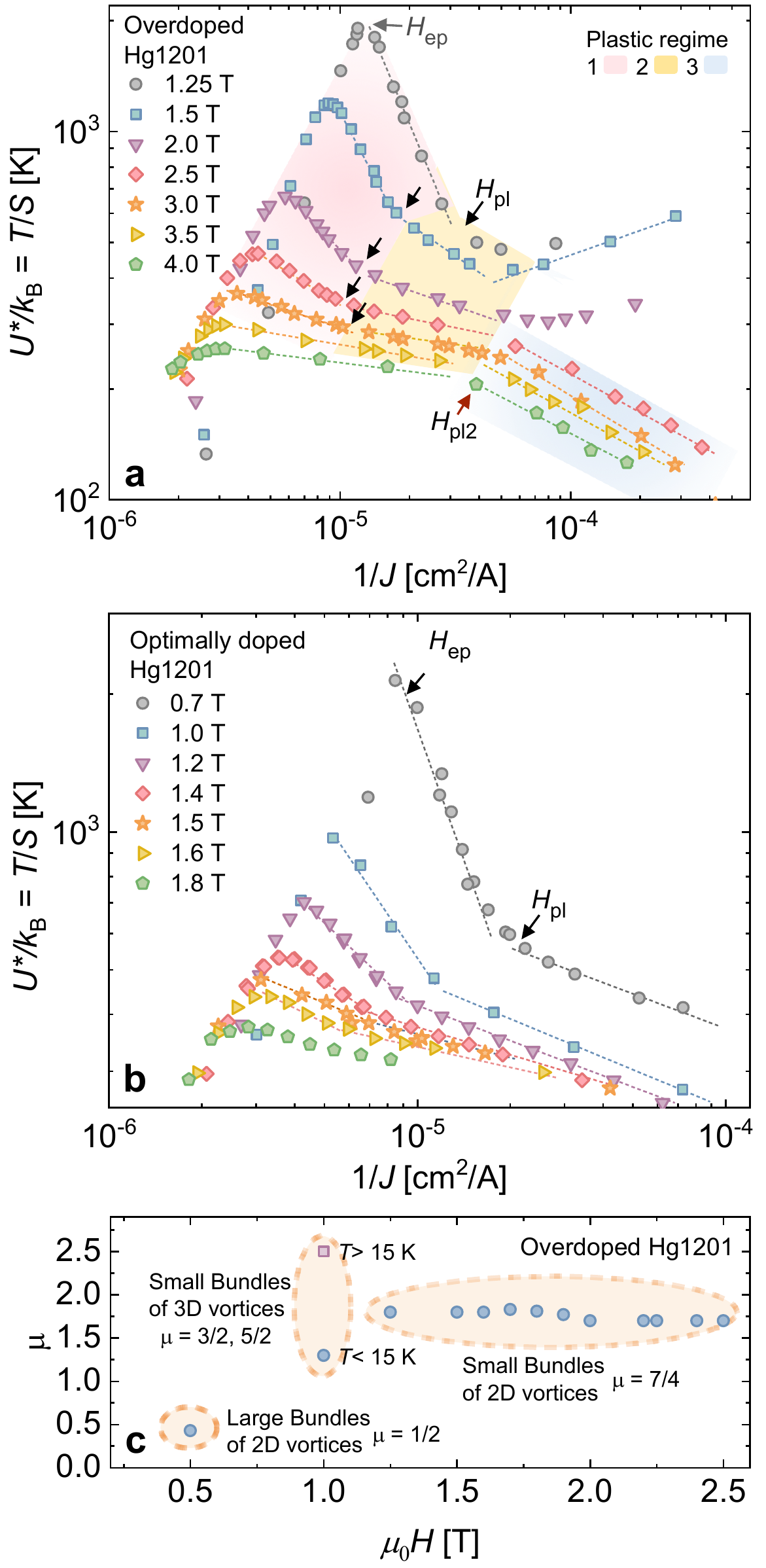}
\caption{\label{fig:FigUvs1J} (a) Energy scale $U^*$ plotted against $1/J$ for the overdoped Hg1201 crystal measured in applied magnetic fields \SIrange{1.25}{4}{\tesla}. Data was collected at \SI{0.2}{\tesla} intervals; for clarity, only select curves are displayed. Dashed lines are examples of linear fits used to extract glassy and plastic exponents, noted in Fig. \ref{fig:FigPhaseDiagram} and the main text. Change from a positive to negative slope suggests a crossover from elastic to plastic vortex dynamics at $H_{ep}$. Data collected at fields below \SI{3.2}{\tesla} exhibit 2 kinks in the plastic flow regime at $H_{pl}$ (lowest $T$ kink, identified with black arrows) and $H_{pl2}$.  Higher fields display only one kink, at $H_{pl2}$ (example labeled with red arrow). Plastic flow regimes 1 ($H_{ep} < H < H_{pl}$), 2 ($H_{pl1} < H < H_{pl2}$), and 3 ($H > H_{pl2}$) identified by red, yellow, and blue shading, respectively. 
(b) $U^*$ versus $1/J$ for the optimally doped Hg1201 crystal. (c) Field dependence of the glassy exponent in the overdoped sample. Labels indicate theoretical expectations that most closely match experimental values.}
\end{figure}

\noindent Accordingly, as we see from Eq.~[\ref{eq:UstarJ}], the exponent can easily be extracted from the slopes of $U^\star$ vs $1/J$ on $\log-\log$ plot. Figure \ref{fig:FigUvs1J} shows $U^\star(1/J)$ collected at multiple fields for both crystals.   Notice that the curves are approximately linear within distinct regimes and exhibit the prominent change from positive to negative slope that is associated with an elastic-to-plastic crossover at $H_{ep}(T)$.  

First focusing on the elastic dynamics in the overdoped crystal, Fig. \ref{fig:FigUvs1J}(c) displays the field dependence of the glassy exponent $\mu$ extracted from linear fits to the elastic regime.  From the \SI{0.5}{\tesla} data, we extract $\mu \approx 0.5$, which matches the expectation for collective creep of large bundles of 2D pancake vortices.  The presence of large bundles in small fields is suggestive of a clean pinning landscape in which long-range $1/r$ vortex-vortex interactions are only weakly perturbed by vortex-defect interactions. When the field is increased to $\SI{1}{\tesla}$, $\mu$ becomes 1.3 at low temperatures $T < \SI{15}{\kelvin}$ and 2.6 at higher temperatures, close to the expectations of $3/2$ and $5/2$ for creep of small bundles of 3D vortices \cite{narlikar2005hts}. At even higher magnetic fields of \SIrange{1}{2.5}{\tesla}, we find that $\mu \approx$ 1.7-1.8, close to the expectation of $7/4$ for creep of small bundles of 2D vortices. 

This change from 2D to 3D to 2D dynamics with increasing field (at low fields) is consistent with predictions from numerical simulations of magnetically interacting vortices in highly anisotropic superconductors, considering long-range nonlinear interactions along the c-axis \cite{CJOlson2003}. Specifically, Ref. \cite{CJOlson2003} found that, at very low fields, vortices are disordered within the planes, uncorrelated in the z-directions, and interact strongly with pinning centers, resulting in relatively high $J_c$.  As the field is increased, a 3D vortex lattice forms as vortices between planes align, painting a picture of 3D vortex lines that are weakly-coupled to an energy landscape of point defects: the lattice stiffness increases, weakening the effectiveness of pinning and effectuating a decrease in $J_c$. The simulations proceed to predict that, at intermediate applied magnetic fields, pancake vortices between planes start to decouple, enabling higher effective pinning therefore higher $J_c$.  Consequently, the 2D-3D-2D transition leads to a dip in $J_c$ such that the simulations are not only consistent with our extracted glassy exponent, but also with the dip in $M(H) \propto J_c(H)$ at $H_{on}$ present in the magnetization loops in Fig. \ref{fig:FigMvsH}.

Notice that the data suggests that large bundles exist at \SI{0.5}{\tesla} whereas small bundles exist at higher fields $> \SI{1.25}{\tesla}$. In many systems, the bundle size increases with increasing field \cite{Blatter1994b}. However, this behavior is consistent with our suggestion in Ref. \cite{Eley2020} that as $H$ increases, the strength of pinning suddenly increases around $H_{on}$ causing the lattice to become more entangled, the bundle size to decrease, and we see both $J_c$ and $\mu$ increase.  Results for the optimally doped crystal are similar, as presented in the phase diagram in Fig. \ref{fig:FigPhaseDiagram} and discussed in detail in Ref. \cite{Eley2020}.  



\subsection{Plastic deformations of the vortex lattice}

Collective creep theory considers elastic deformations of the vortex lattice and neglects dislocations that may govern vortex dynamics in the plastic regime. In this regime, the elastic pinning barrier is sufficiently high such that plastic deformations of the vortex lattice are more energetically favorable. Abulafia et al. \cite{Abulafia1996} first suggested that a  \textit{non-diverging} (as $J \rightarrow 0$) energy barrier $U(J)$  may be suggestive of plastic creep, compared to the \textit{diverging} elastic barrier described in Eq. \eqref{eq:UstarJ}. Specifically, they applied an expression from dislocation theory, replaced strain with $J_c$, and found that the plastic activation barrier $U_{pl}(J) = U_{pl}^0(1-J^{1/2}/J_0)$, where $J_0$ is related to the plastic critical current density.  They further showed that this described data collected on a YBa$_2$Cu$_3$O$_{7-x}$ crystals, in a regime in which there was other evidence of plastic dynamics, such as a decrease in $U$ with magnetic field (whereas an increase is predicted for elastic creep). Consequently, it has become common to identify a change from elastic to plastic creep as a sudden change in slope on a $\log U - \log (1/J)$ plot, identifying the plastic flow regime as $U_{pl} \sim (1/J)^p$, for $p<0$, and many studies have observed a sharp change in slope with (in some cases) strikingly linear behavior on both sides of the transition
\cite{Haberkorn2011b, Rouco_2014, Sun_2013, PhysRevB.81.014503, Sundar2017, Zhou2016a, Miu2012, PhysRevB.86.094527, Sun2015d, Galluzzi_2016, Tahan_2011, PhysRevB.64.220502, Miu2020, Ionescu2018}. At high temperatures and fields $H>H_{ep}$ in our samples, we see that the slope of the $U^*$ versus $1/J$ plot shown in Fig. \ref{fig:FigUvs1J} becomes negative, indicative of plastic flow.

\begin{figure}[ht]
\centering
\includegraphics[height=4.4in]{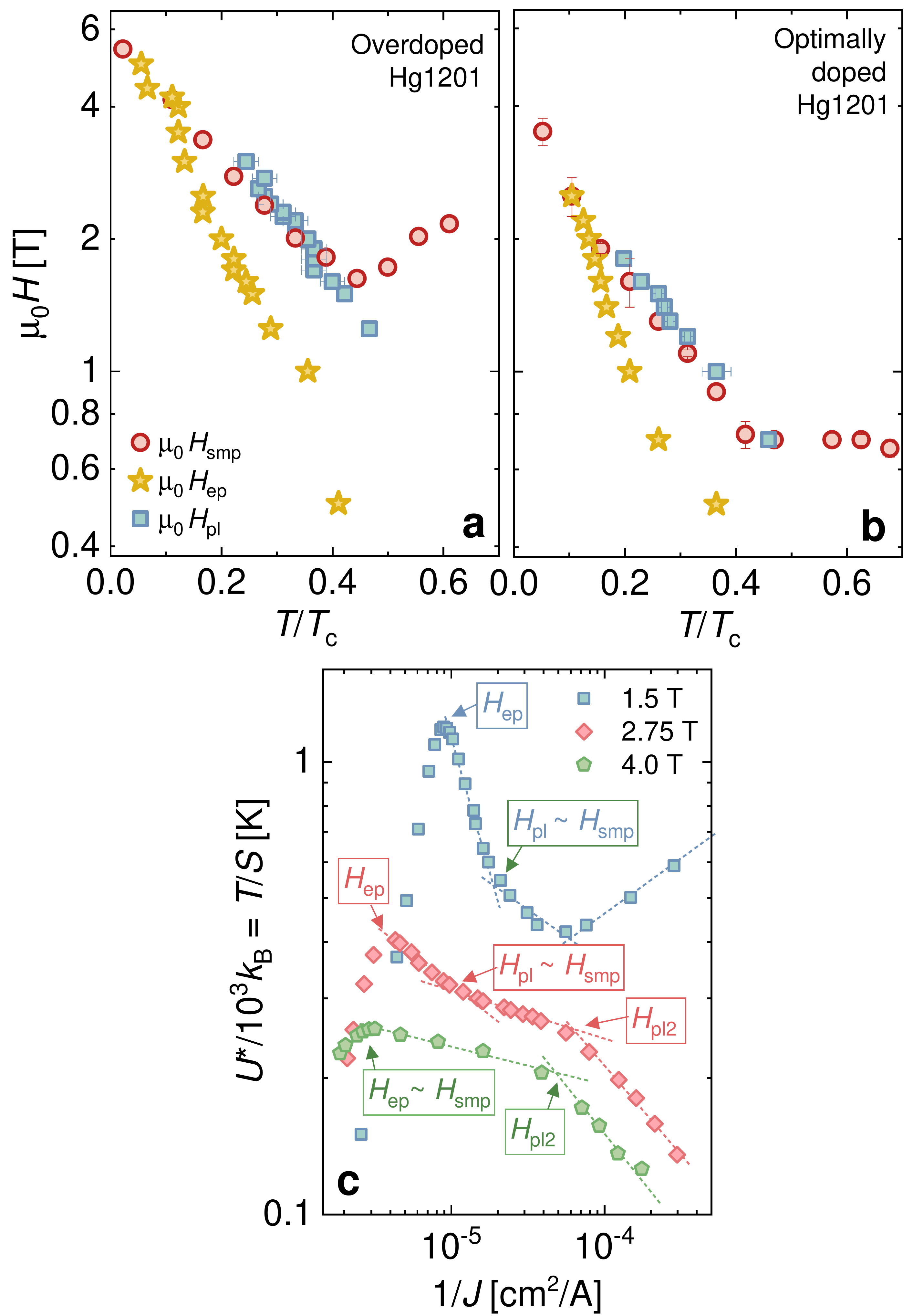}
\caption{\label{fig:FigPDUvs1J} Field - Temperature phase diagram for (a) overdoped and (b) opitmally doped Hg1201 crystals showing that the second magnetization peak coincides with the elastic to plastic crossover only at high fields and with the kink in the plastic flow regime (at $H_{pl}$) at lower fields.
(c) $U^*$ versus $1/J$ for overdoped Hg1201 crystal showing examples of 3 different types of behavior in plastic regime: low fields show two distinct negative slopes then an anomalous transition to a positive slope at low $J$; intermediate fields display 3 distinct negative slopes; high fields exhibit 2 distinct negative slopes. }
\end{figure}

One or two kinks are clearly evident within the plastic flow regime, at the junction between linear regions with distinct slopes.   For applied magnetic fields below approximately \SI{1}{\tesla} in the overdoped sample (Fig. \ref{fig:FigUvs1J}a), we either do not observe an elastic-to-plastic transition or capture little of the plastic regime within the temperature range of our measurements. At higher magnetic fields, we observe 3 different trends in the $U(1/J)$ data, exemplified in Fig. \ref{fig:FigPDUvs1J}(c). First, for fields between \SI{1.25}{\tesla} and \SI{2}{\tesla}, two kinks are clearly evident within the plastic flow regime. We label the kink at the lowest temperature (lowest $1/J$) as $H_{pl}$; the kink that appears at a higher temperature (higher $1/J$) is followed by a transition to a positively sloped region that corresponds to the anomalous non-monotonicity in $S(T)$ around \SI{45}{\kelvin}, seen in Fig. \ref{fig:FigCreep}(a). Second, data collected in fields between \SI{2}{\tesla} and \SI{3}{\tesla} also shows two kinks defining 3 regions with distinct, negative slopes: we again label the lowest temperature kink as $H_{pl}$ and also label the higher temperature one as $H_{pl2}$. Last, for magnetic fields higher than approximately \SI{3}{\tesla}, there appears to be only two distinct negatively sloped regions: here, we label the junction as $H_{pl2}$. 

In the following discussion, we refer to the different regions as plastic regime 1 ($H_{ep} < H < H_{pl}$), regime 2 ($H_{pl} < H < H_{pl2}$), and regime 3 ($H > H_{pl2}$), identified by the red, yellow, and blue shaded regions in Fig. \ref{fig:FigUvs1J}. Many questions arise from the data in Fig. \ref{fig:FigUvs1J}(a). First, is one of these transitions in the plastic flow regime responsible for the second magnetization peak? What type of plastic dynamical arrangements occur in plastic regimes 1, 2, and 3? What causes the sudden increase in $U^*$ around \SI{46}{\kelvin} for the low-field data?

To answer the first question, we plot a comparison of the temperature dependence of $H_{ep}$, $H_{smp}$, and $H_{pl}$ in
Fig. \ref{fig:FigPDUvs1J}(a,b). The second magnetization peak $H_{smp}$ coincides with $H_{ep}$ at high fields and we see a clear correspondence between $H_{smp}$ and $H_{pl}$ at lower fields.  To our knowledge, the latter correspondence has not been previously identified. Distinguishing the dynamics in plastic regimes 1 and 2 may reveal the origin of the second magnetization peak.

\begin{figure}[h!]
\centering
\includegraphics[width=3.3in]{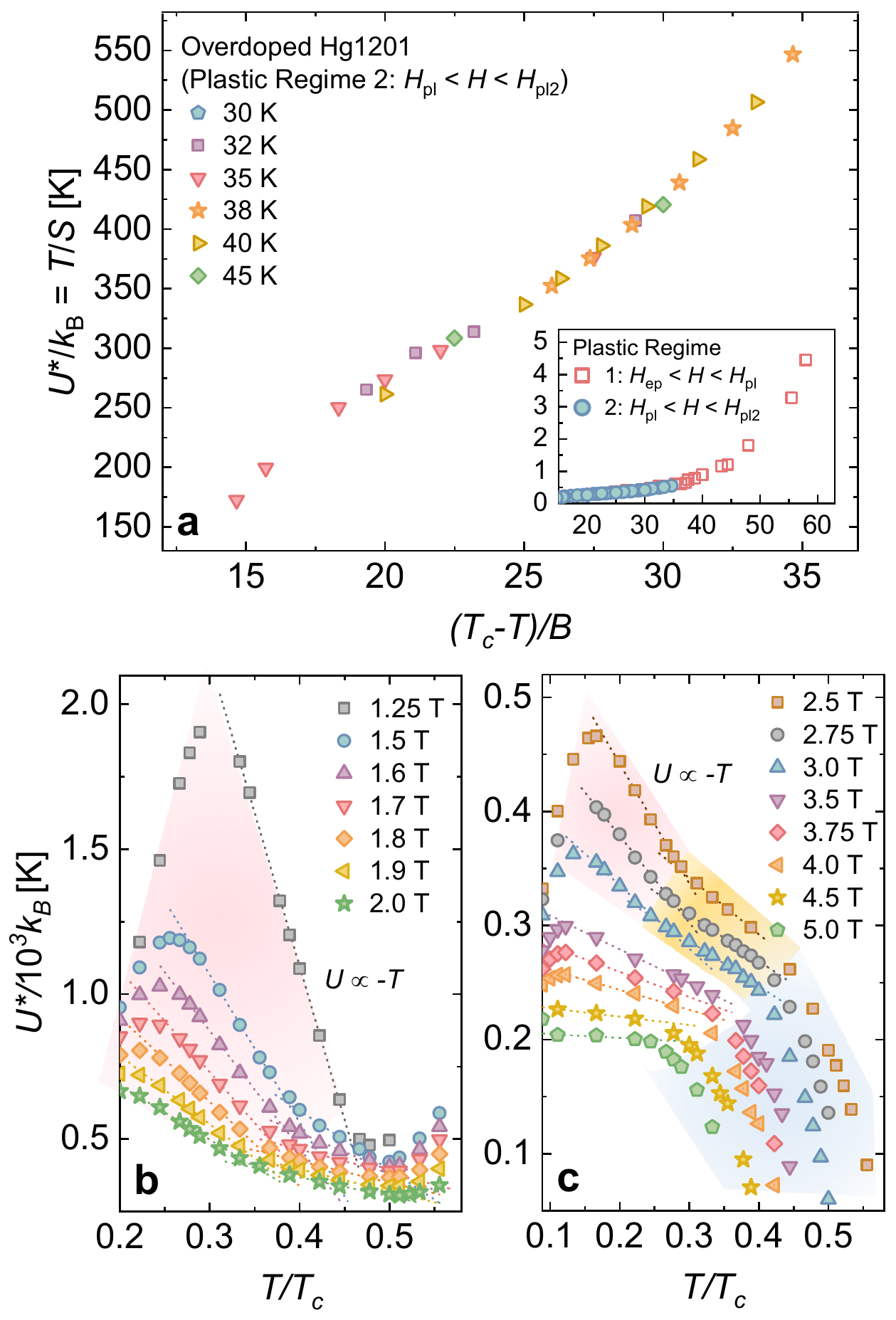}
\caption{\label{fig:FigUvsBvsT}  (a) Energy scale $U^*$ versus $(T_c-T)/B$ for overdoped Hg1201 sample showing that data collapses onto a single curve such that $U^* \sim 1/B$ in plastic regime 2. Inset shows collapse onto single curve for regimes 1 and 2. (a,b) $U^*$ versus temperature, normalized to $T_c$, at (b) low fields and (c) higher fields. Plastic flow regimes 1, 2, and 3 identified by red, yellow, and blue shading, respectively. It is unclear whether the behavior in the unshaded region in (c) captures dynamics similar to regimes 1 or 2. Dashed lines are linear fits.} 
\end{figure}

Plastic deformations of the vortex lattice bend/displace vortices on a scale $u \sim a_0$, for mean intervortex separation $a_0 \sim \sqrt{\Phi_0/B}$ and flux quantum $\Phi_0$, such that the expected field-dependence of the plastic energy barrier is \cite{PhysRevLett.65.259, Blatter1994b}
\begin{equation}\label{eq:EqUpl}
U_{pl} \sim \varepsilon \varepsilon_0 a_0 \propto (T_c-T)/B^\alpha.
\end{equation}
Here, $\varepsilon_0 =\Phi_0^2/(4\pi\mu_0\lambda_{ab}^2)$ is the vortex line tension, $\varepsilon = \sqrt{m_{ab}/m_c}$ is the electronic mass anisotropy (for $m_{ab}$ and $m_c$ are the masses in the ab-plane and along the c-axis, respectively). Hence, we now look at the temperature and field dependence of the plastic energy barrier $U_{pl}(H,T)$, presented in Fig. \ref{fig:FigUvsBvsT}.

Figure \ref{fig:FigUvsBvsT} shows that the plastic activation energy $U_{pl}$ in regime 2 decreases with magnetic field, consistent with plastic dynamics, as opposed to collective creep in which $U_{el}$ increases with magnetic field \cite{Abulafia1996}. Specifically, we find that $U_{pl} \propto (T_c-T)/B$ for the overdoped crystal. Notice that the data collected at different temperatures all collapse onto a single curve and appear roughly linear in the figure. Though plastic creep theory predicts $\alpha \sim 1/2$ \cite{Kierfeld2000, Geshkenbein1989}, faster than $1/\sqrt{B}$ behavior has been previously observed \cite{Abulafia1996, Choi2017, WU201639, Wang2017, Obolenskii2000} and associated with entangled vortex liquid behavior from \textit{stronger} pinning due to point disorder \cite{PhysRevLett.80.1070} than for $\alpha \sim 1/2$.  The energy $U_{pl}$ decreases faster with increasing field because the entangled vortices become cut and disconnected, \cite{PhysRevLett.80.1070, PhysRevLett.65.259} therefore moving faster (faster $S$).

Plotting $U_{pl}$ versus $(T_c-T)/B$ for plastic regimes 1 and 2 on the same plot, shown in the inset to Fig. \ref{fig:FigUvsBvsT}(a), we see that all data collapse onto the same curve. It is evident that $U_{pl}$ rises far more rapidly with increasing $(T_c-T)/B$ than in regime 2 as plastic regime 1 includes lower fields than regime 2. Data for the plastic activation barrier for regime 3 (not shown in the inset) deviates from this trend.

Equation \ref{eq:EqUpl} indicates that, at fixed magnetic fields, the plastic activation barrier should decrease linearly with temperature \cite{PhysRevLett.65.259}. This behavior is evident in multiple regions of the $U(T/T_c)$ plot in Fig. \ref{fig:FigUvsBvsT}(b,c), disregarding the low-field high-$T/T_c > 0.5$ data that corresponds to the anomalous peak in $S(T)$.
Also, note that at high fields $\mu_0H \geq \SI{3.5}{\tesla}$ and low temperatures $T/T_c < 0.3$, $U_{pl}$ becomes relatively insensitive to temperature.

\begin{figure*}[t!]
\centering
\includegraphics[width=6in]{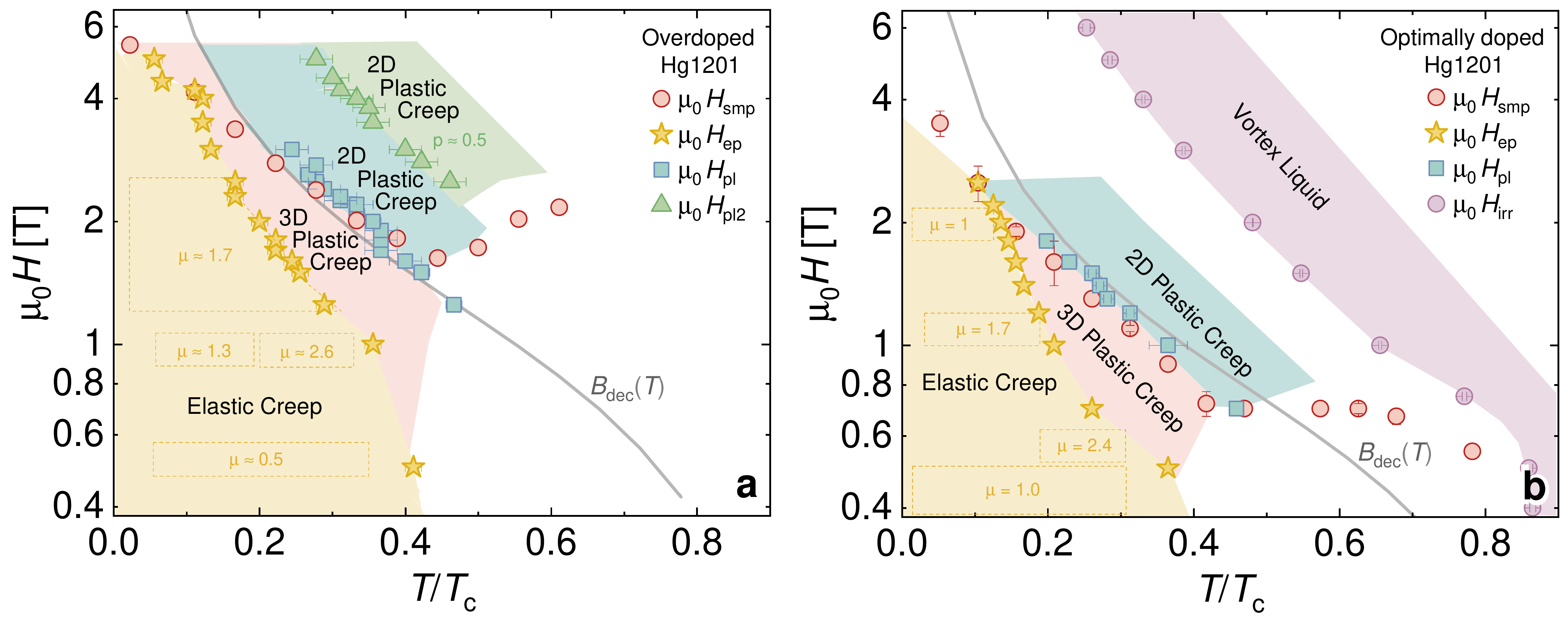}
\caption{\label{fig:FigPhaseDiagram} Vortex phase diagram determined from creep measurements overlaid with position of second magnetization peak at $H_{smp}$ for (a) overdoped and (b) optimally doped Hg1201 crystals. The gray curve is the prediction for the 3D-to-2D decoupling field based on Eq. \ref{eq:EqBdec}. Overlap between $B_{dec}$ suggest the second magnetization peak originates from a 3D-to-2D dimensional crossover from 3D-to-2D plastic dynamics.}
\end{figure*}

We now evaluate the possibility of a 3D-to-2D dimensional crossover causing the kinks in the plastic flow region.  In highly anisotropic cuprate superconductors, the vortex lattice is composed of pancake vortices that form within the CuO$_2$ layers and interact both magnetically and through Josephson coupling between the layers. In the 3D regime, significant interlayer coupling causes vortices to behave as chains, somewhat aligned along the c-axis, experiencing 3D-like fluctuations, resembling lines in isotropic superconductors.  A couple mechanisms may induce quasi-2D behavior, in which the pancakes move independently within the plane:  the strength of interactions between pancake vortices within a layer may surpass that of coupling between adjacent layers \cite{Vinokur1990} or destruction of phase coherence by fluctuations may ultimately lead to Josephson decoupling \cite{Deutscher1990}. Changing the magnetic field can induce this transition partially because an increase in the magnetic field reduces the intralayer vortex separation whereas the distance between planes remains fixed. A 3D-to-2D transition may manifest as a sudden change in magnetization data: when vortices become decoupled between the layers, pancakes vortices within a layer can more freely re-position themselves to minimize their energy, fully exploiting available pinning centers, resulting in an increase in $J_c$ (or slower decrease with field) \cite{PhysRevB.49.12984}.

Let us now consider the expected decoupling field $B_{dec}(T)$ based on Josephson decoupling arising from destruction of phase coherence \cite{Hardy1994}:
\begin{equation}\label{eq:EqBdec}
    B_{dec} = \Phi_0^3[\ln(\lambda_{ab}/s)+1.12   ]/4\pi\mu_0\lambda_{ab}^3\gamma^2k_BT,
\end{equation}
\noindent which considers the Lawrence-Doniach expression for line tension \cite{Clem_1992}. To calculate $B_{dec}(T)$ in our overdoped crystal, we know that the distance between CuO$_2$ planes $s = \SI{0.952}{\nano\meter}$ \cite{PhysRevB.59.7209}, $\gamma = 25$ \cite{Hofer1998}, and $\lambda_{ab}(0) = \SI{156}{\nano\meter}$ \cite{Hofer1998}. In the case of the optimally doped crystal, we use $\lambda_{ab}(0) = \SI{154}{\nano\meter}$, $\gamma = 32$ \cite{Eley2020}, and $s = \SI{0.953}{\nano\meter}$ \cite{PhysRevB.59.7209}. For the temperature dependence of the penetration depth, we apply the two-fluid expression $\lambda_{ab}(T)=\lambda_{ab}(0)[1-(T/T_c)^4]^{-1/2}$, which was found to accurately describe $\lambda_{ab}(T)$ in Hg1201 \cite{Hofer1998}. Based on these parameters, we plot the calculated $B_{dec}(T)$ in the phase diagram in Fig. \ref{fig:FigPhaseDiagram}. Remarkably, we find striking overlap between the predicted $B_{dec}(T)$, $H_{pl}(T)$, and $H_{smp}(T)$ for both samples.  This suggests that the second magnetization peak and crossovers in the plastic flow regime at $H_{pl}$ originate from a dimensional crossover in the plastic flow regime. Thermal fluctuations can cause reduced instantaneous intralayer vortex separation. This may nucleate a decoupling transition to 2D dynamics, such that the 3D to 2D transition occurs at lower fields with increasing temperature \cite{CJOlson2003}, congruous with the SMP and plastic flow transitions delineated in the phase diagram in Fig. \ref{fig:FigPhaseDiagram}.

Lastly, we seek to understand dynamics in plastic regime 3. As the field is further increased, stronger interlayer vortex-vortex interactions stiffen the lattice, weakening vortex-defect interactions engendering a more rapid decrease in $J_c$ \cite{CJOlson2003}. In plastic regime 3, we extract a plastic exponent $|p| \sim 0.4-0.5$ [slope of $\log U^* - \log (1/J)$] from the data in Fig. \ref{fig:FigUvs1J}(b) that is similar at all fields, and consistent with the expectation of dislocation mediated motion of vortices \cite{Abulafia1996}, though it is unclear what the exponent should be for 2D plastic creep.


In this work, we show evidence of a dimensional crossover in the plastic flow regime. Yet there remains transitions that we have identified to regimes in which the dynamics arrangements are unclear. Further work is warranted, with a particular interest in neutron scattering studies, which could help reveal these phases and clarify the mechanism behind the unexplained transitions.  For example, such studies would help evaluate whether the vortex lattice may undergo a structural transition.  These results may also shed light on understanding plasticity in other systems, including materials containing skyrmions \cite{OLSONREICHHARDT201452, Reichhardt_2016, PhysRevLett.114.217202}, domain walls \cite{PhysRevLett.80.849, Lindquist2015}, and charge density waves \cite{Isakovic2006, Brazovskii2004}.





\section{Methods}

\subsection{Growth and Structural Characterization}
Our study includes results from two HgBa$_2$CuO$_{4+x}$ (Hg1201) single crystals: an optimally-doped crystal of dimensions $1.28 \times 0.84 \times 0.24$ mm$^3$ and an overdoped crystal of dimensions \SI[product-units = single]{1.5 \pm 0.05 x 1.05 \pm 0.05 x 0.23}{\cubic\milli\metre}. The thickness of the overdoped sample was found by calculating the superconducting volume $V = -4\pi(1-D)\frac{dm}{dH}$ from a measurement of the Meissner slope $\frac{dm}{dH}$, using the demagnetization factor \cite{Pardo2004} $D=0.65$, and dividing by the measured lateral dimensions.  The anisotropy $\gamma$ of the samples is 32 (optimally-doped \cite{Eley2020}) and 25 (overdoped \cite{Hofer1998}).

Both samples were grown at Los Alamos National Laboratory using an encapsulated self-flux method described in Ref.~[\onlinecite{zhao06}]. The optimally doped crystal was subsequently heat-treated at 350$^o$C in air and quenched to room temperature \cite{yamamoto00}, whereas the the overdoped crystal was heat-treated in approximately 2 bar O$_2$ at 300$^o$C. 

\subsection{Magnetometry}
Magnetization measurements $M(T,H,t)$ were performed using a Quantum Design superconducting quantum interference device (SQUID) magnetometer, in which the magnetic field was aligned with the sample's c-axis ($ H \parallel c$).  The data for the overdoped crystal was collected using a QD MPMS3 system, whereas the optimally doped crystal was measured using a QD MPMS XL: this difference accounts for the lower point densities for the latter due to significantly slower measurements (slower field sweep rate).

When collecting magnetization loops $M(H)$, the field was first swept to $\SI{-3}{\tesla}$ or $\SI{-4}{\tesla}$ to establish the critical state (full flux penetration throughout the sample).  The lower branch of the loop was subsequently measured as the field was ramped from \SI{0}{\tesla} to $\SI{7}{\tesla}$, and the upper branch was collected as the field was swept down to $\SI{-7}{\tesla}$, then ramped back up to zero.

Creep data were taken using the standard approach \cite{Yeshurun1996b}.  First, we establish the critical state sweeping the field high enough that vortices, which first enter at the sample peripheries, permeate the center of the sample and the Bean critical state model defines the vortex distribution.  Following standard practice, this requires a sweep of $\Delta H > 4H^\star$ (for $H^\star$ is the field of full flux penetration) and verification that the initial $M(T,H)$ lies on the magnetization loop. Subsequently, we capture the decay in the magnetization $M(t)$ by repeatedly measuring $M$ every $\sim$\SI{10}{\second} at a fixed $T$ and $H$: we measure $m(t)$ for an hour in the upper branch, preceded by a brief measurement in the lower branch that enables determining the background (e.g. from sample holders). After subtracting this background (average of the upper and lower branches) and adjusting the time to account for the difference between the initial application of the magnetic field and the first measurement, $S \equiv d \ln m / d \ln t$ was determined from the slope of a linear fit to $\ln m$ versus $\ln t$. 

\section*{Author Contributions}
S. E. conceived and designed the experiment. S. E. and H. M. C. collected and analyzed the magnetization data. M. B. V. wrote software to perform data analysis. M. K. C. and E. D. B. grew the samples. B. G. performed transmission electron microscopy and analyzed the microstructural data. S.E. wrote the manuscript. All authors commented on the manuscript.

\section*{Data availability}
All data presented in this work are available from the corresponding authors upon reasonable request.

\section*{Acknowledgements}
We would like to thank V. Vinokur, C. J. O. Reichhardt, and C. Reichhardt for useful discussions regarding the plastic flow regime. This material is based upon work supported by the National Science Foundation under Grant No. DMR-1905909 (magnetization measurements, data analysis, and manuscript composition). M.K.C. acknowledges support from the U.S. Department of Energy, "Science at 100 T" (crystal synthesis).  E.D.B. acknowledges support from the U.S. Department of Energy (DOE), Office of Basic Energy Sciences, Division of Materials Science and Engineering under project “Quantum Fluctuations in Narrow-Band Systems."

\bibliography{apssamp}

\end{document}